\documentclass[12pt]{article}
\usepackage[cp1251]{inputenc}
\usepackage{cyr}
\usepackage{epsf}
\usepackage{pazh}
\usepackage[russian]{babel}
\usepackage{morefloats}
\tightenlines

\def\*{$^{*}$}
\def\Б{$^{\mbox{\small Б}}$}
\def\В{$^{\mbox{\small В}}$}
\def\Ч{$^{\mbox{\small Ч}}$}
\def\З{$^{\mbox{\small З}}$}
\def\Д{$^{\mbox{\small Д}}$}
\def\ЕТЗУ{ЬТЗ~У$^{-1}$}
\def\ЕТЗУН{ЬТЗ~УН$^{-2}$~У$^{-1}$}

\sloppypar

\begin{document}
\baselineskip 21pt

\vspace{52mm}

\title{\bf Precursors of Short Gamma-Ray Bursts in the SPI-ACS/INTEGRAL Experiment}

\vspace{2mm}

\author{\bf \hspace{-1.3cm}  \ \
P. Yu. Minaev\affilmark{1*}, A. S. Pozanenko\affilmark{1,2}}
\vspace{2mm}
\affil{
{\it $^1$ Space Research Institute, Russian Academy of Sciences, Moscow, Russia}\\
{\it $^2$ MEPhI National Research Nuclear University, Moscow, Russia}
}

\vspace{2mm}

\sloppypar \vspace{12mm} \noindent
We have analyzed the light curves of 519 short gamma-ray bursts (GRBs) detected in the SPI-ACS/INTEGRAL experiment from December 2002 to May 2014 to search for precursors. Both the light curves of 519 individual events and the averaged light curve of 372 brightest bursts have been analyzed. In a few cases, we have found and thoroughly studied precursor candidates based on SPI-ACS/INTEGRAL, GBM/Fermi, and LAT/Fermi data. A statistical analysis of the averaged light curve for the entire sample of short bursts has revealed no regular precursor. Upper limits for the relative intensity of precursors have been estimated. No convincing evidence for the existence of precursors of short GRBs has been found. We show that the fraction of short GRBs with precursors is less than 0.4\% of all short bursts.

\vspace{5mm}



\vfill
\noindent\rule{8cm}{1pt}\\
{$^*$ E-mail: $<$minaevp@mail.ru$>$}

\clearpage

\section*{INTRODUCTION}
\noindent

Bimodality in the gamma-ray burst (GRB) duration distribution was discovered in a series of KONUS experiments (Mazets et al. 1981) and was subsequently confirmed by more extensive statistical data from the BATSE experiment (Kouveliotou et al. 1993), where a robust GRB duration parameter Т$_{90}$ was proposed. GRBs with a duration Т$_{90}$ < 2 s are deemed short. The characteristic duration separating the long bursts from the short ones depends on the spectral range (see, e.g.,Minaev et al. 2010b). A harder energy spectrum and a less distinct spectral lag are typical for short GRBs (Kouveliotou et al. 1993; Norris et al. 2005). The phenomenological properties of short GRBs are considered most comprehensively in Donaghy et al. (2006) and Berger (2014).

The class of short GRBs is believed to be associated with the merger of compact components (neutron stars, black holes) in a binary system (Paczynski 1986; Meszaros and Rees 1992, 1997; Rosswog and Ramirez-Ruiz 2003a; Rosswog et al. 2003b). This is confirmed by the absence of observational signatures of a supernova in the optical afterglow light curves for short GRBs and by the localization of the sources of short GRBs in galaxies of both late and early types with a low star formation rate (Berger 2014). However, in one case (GRB 130603B), a signature of a kilonova was detected in the light curve (Tanvir et al. 2013). The predicted (Li and Paczynski 1998) kilonova is an optical and infrared transient produced by the ejection of radioactive matter during the merger of a binary system where one of the components is a neutron star.

Activity before the beginning of the main episode (precursor) was detected in the light curves of some GRBs (Koshut et al. 1995; Lazzati 2005; Troja et al. 2010). So far there is no single definition of precursor. For example, in Troja et al. (2010) the less intense and shorter burst activity episode preceding the main one is considered to be a precursor. In Koshut et al. (1995) an additional condition is imposed on the precursor properties: the time interval between the precursor and the main episode of a GRB must exceed the duration Т$_{90}$ of the main episode. As a rule, the GRB light curves have a complex structure and consist of several pulses, both overlapping between themselves and well separated in time, with the intensity and duration of the pulses as well as the time interval between them being independent of their relative positions in the GRB light curve (see, e.g., Mitrofanov et al. 1998). Therefore, the possibility that the detected precursor candidate is actually an individual pulse of the main episode and is not associated with a different emission mechanism and/or source must not be ruled out either. The probability of this can be significant for the precursors detected by Troja et al. (2010). Therefore, the positive results of the search for the precursors of short GRBs obtained previously by Koshut et al. (1995) and Troja et al. (2010) require an additional verification.

The precursors of long bursts can be explained in terms of the main models of their sources and may be associated with the shock breakout through the surface of the GRB progenitor star (see, e.g., MacFadyen and Woosley 1999). The precursors of short bursts are not predicted within the main models, which also casts doubt on their existence. There is an assumption that they can be associated with the reconnection of magnetic fields lines of neutron stars before their merger (Troja et al. 2010).

This paper is devoted to searching for precursors in the individual light curves and the averaged light curve of short GRBs detected in the SPI-ACS/INTEGRAL experiment and to a detailed study of the precursor candidates found for a few GRBs.

\section*{ANALYSIS OF THE LIGHT CURVES FOR SHORT GAMMA-RAY BURSTS DETECTED BY SPI-ACS/INTEGRAL}
\subsection*{The Anticoincidence Shield of the SPI Spectrometer (SPI-ACS)} \noindent

The INTEGRAL observatory was launched into a highly elliptical orbit (the perigee and apogee of its initial orbit are 9000 and 153 000 km, respectively) with a period of 72 h on October 17, 2002 (Jensen et al. 2003). The observatory consists of the IBIS, SPI, JEM-X, OMC telescopes and the SPI anticoincidence shield (SPI-ACS). All of the aperture telescopes (SPI, IBIS, JEM-X, OMC) onboard the observatory are aligned.

SPI-ACS composed of 91 bismuth germanate (BGO) crystals with a maximum effective area of about 0.3 m$^{2}$ (von Kienlin et al. 2003) is used to reduce the background of the SPI detectors associated with the interaction of the equipment with cosmic rays. Each BGO crystal is viewed by two photomultipliers (PMTs), and the counts from all PMTs are recorded in a single channel. SPI-ACS records photons from almost all directions. The direction opposite to the SPI field of view is least sensitive (see, e.g., Vigano and Mereghetti 2009). SPI-ACS has a lower sensitivity threshold of $\sim$ 80 keV - the physical properties of the individual BGO assemblies (detector + PMT + discriminator) slightly differ and, therefore, have different lower energy thresholds: from 60 to 120 keV; the upper energy threshold is $\sim$ 10 MeV. The SPI-ACS time resolution is 50 ms (von Kienlin et al. 2003).

IBAS (INTEGRAL Burst Alert System; Mereghetti et al. 2003) is used for the identification of bursts in the SPI-ACS data. The IBAS software algorithm identifies events on nine different time scales (0.05, 0.1, 0.2, 0.4, 0.8, 1, 2, and 5 s) provided that the event statistical significance with respect to the mean background is 9, 6, 9, 6, 9, 6, 9, and 6$\sigma$, respectively. The light curves of the events identified by this algorithm (containing data from -5 to 100 s relative to the trigger time) are publicly accessible (http://isdcarc.unige.ch/arc/FTP/ibas/spiacs/). IBAS is also used in an automatic analysis of the IBIS/ISGRI data, but IBAS is not used simultaneously with the SPI spectrometer.

\subsection*{The Data Processing Algorithm}

We selected a sample of 519 short GRBs detected from December 2002 to May 2014 in the SPI-ACS/INTEGRAL experiment. To produce the sample, we used the master list by K. Hurley (http://www.ssl.berkeley.edu/ipn3/chronological.txt), which is a compilation of known GRB catalogs and, as a consequence, is the most complete sample of GRBs (more than 6500 events).

For each GRB detected since the beginning of the INTEGRAL mission and contained in the master list, we constructed and analyzed the light curve based on SPI-ACS/INTEGRAL data. Thus, all GRBs of the sample were independently confirmed at least by one different experiment. A script (http://isdc.unige.ch/$\sim$savchenk/spiacs-online/spiacs-ipnlc.pl) was used as the source of input SPI-ACS data. In the case of statistically significant (more than six standard deviations) detection of a given GRBs in the SPI-ACS data, we calculated the burst duration $T_{90}$ (for the duration calculation algorithm, see, e.g., Minaev et al. 2010a). The GRBs with a duration of less than 2 s were deemed short and were included in our sample.

The light curves of the investigated short GRBs were aligned relative to the peak in the light curve with a time resolution of 50 ms (the original time resolution of the SPI-ACS experiment) and were investigated in the time interval (-150; +200) s. The data processing procedure is similar to the procedure described in detail in Minaev et al. (2010a) and consists of the following steps:

1) \emph{Fitting the background and subtraction of the background model from the light curve.} Minaev et al. (2010а) investigated the background variations on various time scales and showed the behavior of the background on a time scale of 350 s to be monotonic, which allows a linear model to be used to fit the background in the intervals (-150; -50) and (+100; +200) s relative to the peak in the light curve. It was also established that the typical background signal changed by no more than 0.3\% during the 350-s time interval. In most cases, the linear model describes well the behavior of the background, but, in several cases, the quality of the fit was unsatisfactory, and such events were excluded from the statistical analysis.

2) \emph{Light curve rebinning} - constructing the light curves with time resolutions of 0.1 and 5 s.

3) \emph{Searching for precursor candidates} on a time scale of 0.1 s in the interval (-5, -2) s relative to the main peak in the light curve and on a time scale of 5 s in the interval (-50, -2) s with a statistical significance of more than six standard deviations.
\\ By a precursor candidate we mean the less intense, short burst activity episode preceding the main one that is offset by more than 2 s from it. Such a criterion allows the probability that the precursor candidate is actually one of the pulses of the main short GRB episode to be minimized (for details, see the Discussion section).

4) \emph{Detailed analysis of the detected precursor candidates} based on data from various experiments (SPI-ACS, GBM/Fermi, LAT/Fermi). The spectral–temporal properties of the precursor (its duration,
energy spectrum, variability, the shape and number of pulses in its light curve) were assumed to differ from those of the main burst phase. Therefore, only in the case of statistically significant deviations of the properties of the precursor candidate from those of the main burst phase did we consider the detected precursor candidate to be actually a precursor.

5) \emph{Averaging of the light curves.} When constructing the averaged light curve, we excluded the events with a low statistical significance (less than ten standard deviations). A total of 372 light curves for short bursts were selected for the averaging procedure. The light curves for this sample of GRBs aligned relative to the main peak and processed according to the procedure described above were summed (the counts of the light curves corresponding to the same instant of time in different light curves were summed).

6) \emph{Estimating the statistical significance and calculating the errors.} When a statistical analysis of the SPI-ACS background was performed, the mean sample (empirical) variance of the background
$D_{E}(B)=(1/(N-1))*\sum\limits_{i=1}^N(B_{i}-\overline{B})^{2} \approx 1.57^{2}B$ turned out to be greater than the Poisson variance $D_{P}(B)=B$, which is consistent with the results of other papers (von Kienlin et al. 2003; Ryde 2003; Minaev et al. 2010a). All the subsequent calculations of the errors and the determination of the statistical significance of the signal above the background take this fact into account.
\\ A possible explanation of this phenomenon is as follows. First, there is an instrumental effect associated with the fact that each scintillation bismuth germanate (BGO) crystal of which the SPI-ACS detector consists is viewed by two PMTs. The variance of the total signal collected from the two PMTs will be greater than the Poisson variance, because the signals from the two PMTs are partially correlated with one another, specifically, in the case where the gamma-ray photon that entered the BGO crystal is recorded by both PMTs. Second, the actually observed variances can be even greater, because the SPI-ACS experiment records the photons of almost all variable sources from all directions. The variance of the variable sources can add 10\% (see, e.g., Bisnovatyi-Kogan and Pozanenko 2011). Almost half of this value can be attributed to the bright variable Galactic sources Cygnus X-1 and Scorpio X-1.

\section*{RESULTS}

\begin{table}[t]
\scriptsize
\vspace{6mm} \centering {{\bf Table 1.} \emph{Precursor candidates for short GRBs of the SPI–ACS/INTEGRAL experiment}}\label{meansp}

\vspace{5mm}\begin{tabular}{c|c|c|c|c|c} \hline\hline
GRB	   &	Start time of 	  &	Time relative   & Significance          & off-axis $^{1}$,     & Detection by other spacecraft $^{2}$     \\
       &   main phase,        & to main         & of precursor,         &  deg.                &          \\
       &   UTC               &  phase, s        &  $\sigma$              &          &                  \\\hline

071030 & 08:52:43.75 & 2.5  &  6.3 & - &Suz$^{7}$, Kon$^{8}$ \\
090510 $^{3}$ & 00:22:59.8 & 0.45  &  4.3 & 140.8 $^{4}$ &Fer$^{9}$, Swi$^{10}$, Kon$^{11}$, Agi$^{12}$, Mes$^{13}$, Suz$^{14}$ \\
100717 & 08:55:05.85  & 3.3  &  12.8 & 40.6 $^{5}$ &Fer$^{13}$, Swi$^{13}$, Mes$^{13}$, Agi$^{13}$ \\
130310 & 20:09:40.9 &4.55  &  10.0 & 59.8 $^{6}$ &Fer$^{15}$, Kon$^{16}$, Mes$^{17}$, Suz$^{17}$, Hend$^{17}$ \\\hline

\multicolumn{6}{l}{}\\
\multicolumn{6}{l}{$^{1}$ The angle between the SPI-ACS axis and the direction to the burst source.}\\
\multicolumn{6}{l}{$^{2}$ Kon - KONUS, Suz - Suzaku, Fer - Fermi, Swi - Swift, Agi - Agile, Mes - Messenger, Hend - HEND-Odyssey}\\
\multicolumn{6}{l}{$^{3}$ The precursor candidate from Troja et al. (2010)}\\
\multicolumn{6}{l}{$^{4}$ The burst source was localized in the UVOT/Swift experiment with coordinates }\\
\multicolumn{6}{l}{R.A., Dec = 333.55271, -26.58266; radius = 1.4$^{\prime\prime}$ (Goad et al. 2009)}\\
\multicolumn{6}{l}{$^{5}$ The burst source was localized in the GBM/Fermi experiment with coordinates }\\
\multicolumn{6}{l}{R.A., Dec = 287.06, -0.66; radius = 8.8$^{\circ}$ (von Kienlin et al. 2014)}\\
\multicolumn{6}{l}{$^{6}$ The burst source was localized in the LAT/Fermi experiment with coordinates }\\
\multicolumn{6}{l}{R.A., Dec = 142.34, -17.23; radius = 0.45$^{\circ}$ (Guiriec et al. 2013)}\\
\multicolumn{6}{l}{$^{7}$ The WAM/Suzaku GRB catalog http://www.astro.isas.jaxa.jp/suzaku/HXD-WAM/WAM-GRB/}\\
\multicolumn{6}{l}{$^{8}$ The KONUS/Wind GRB catalog http://gcn.gsfc.nasa.gov/konus\_grbs.html}\\
\multicolumn{6}{l}{$^{9}$ (Guiriec et al. 2009; Ohno and Pelassa 2009). $^{10}$ (Hoversten et al. 2009). $^{11}$ (Golenetskii et al. 2009). }\\
\multicolumn{6}{l}{$^{12}$ (Longo et al. 2009). $^{13}$ (Hurley 2014). $^{14}$ (Ohmori et al. 2009). $^{15}$ (Guiriec et al. 2013; Xiong and Chaplin 2013). }\\
\multicolumn{6}{l}{$^{16}$ (Golenetskii et al. 2013a). $^{17}$ (Golenetskii et al. 2013b).}\\

\end{tabular}
\end{table}

Precursor candidates were found in the individual light curves of GRB 071030, GRB 100717, and GRB 130310 on a time scale of 0.1 s with a statistical significance of more than six standard deviations (Figs. 1–3, respectively; Table 1). The candidates can be associated not only with the precursors but also with the background fluctuations, other GRBs, and the interaction of charged particles with the detectors. Therefore, first of all it is necessary to test the candidates found in the SPI-ACS data for reliability.

\begin{figure}[h]
\epsfxsize=17cm \hspace{0cm}\epsffile{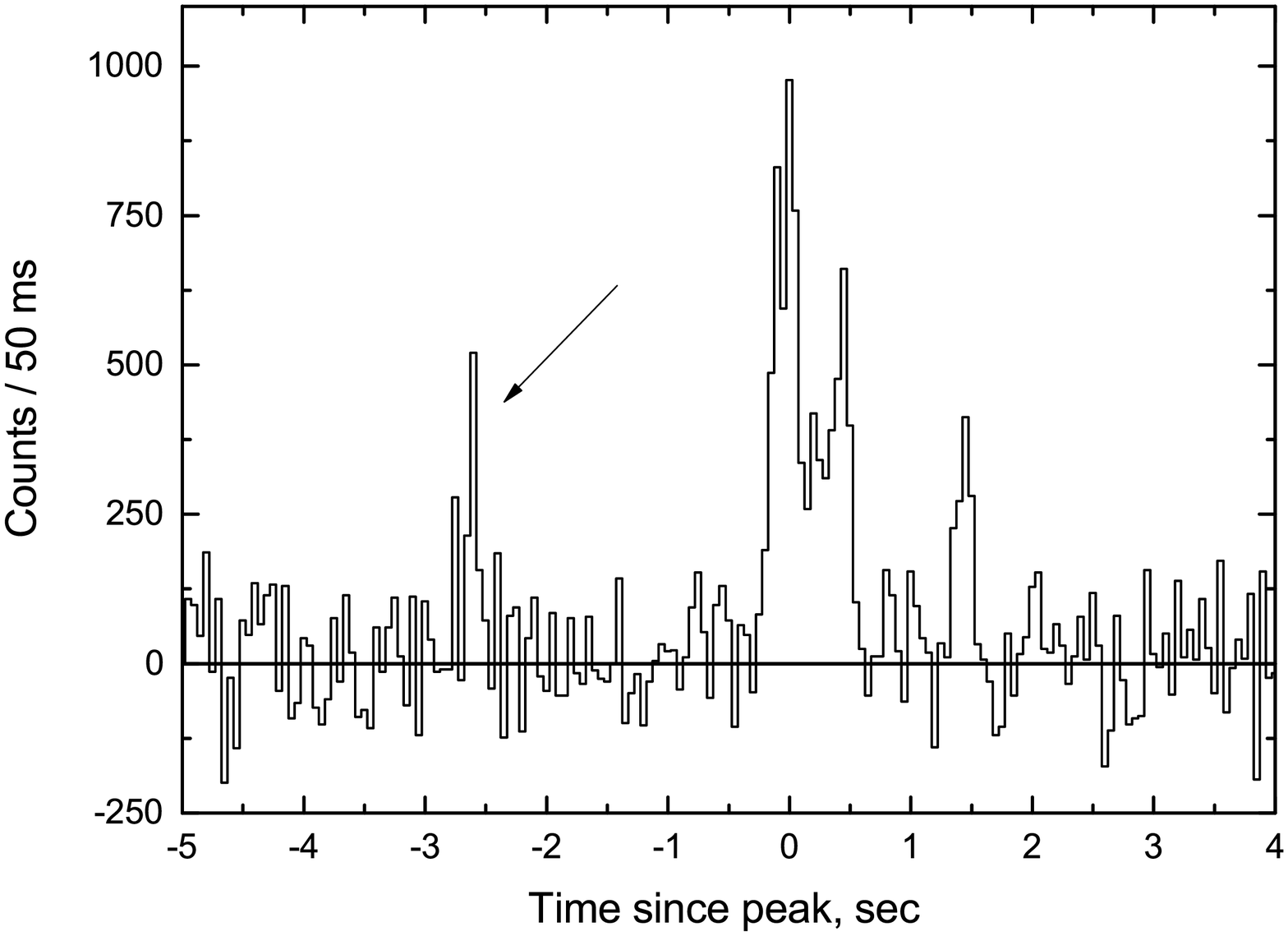}
\\ \textbf{Fig. 1.} \emph{Light curve of GRB 071030 from the SPI-ACS/INTEGRAL data. The time relative to the peak in seconds is along the horizontal axis. The number of counts in 0.05 s is along the vertical axis. The arrow marks the precursor candidate.}
\end{figure}

\begin{figure}[h]
\epsfxsize=18cm \hspace{-0.7cm}\epsffile{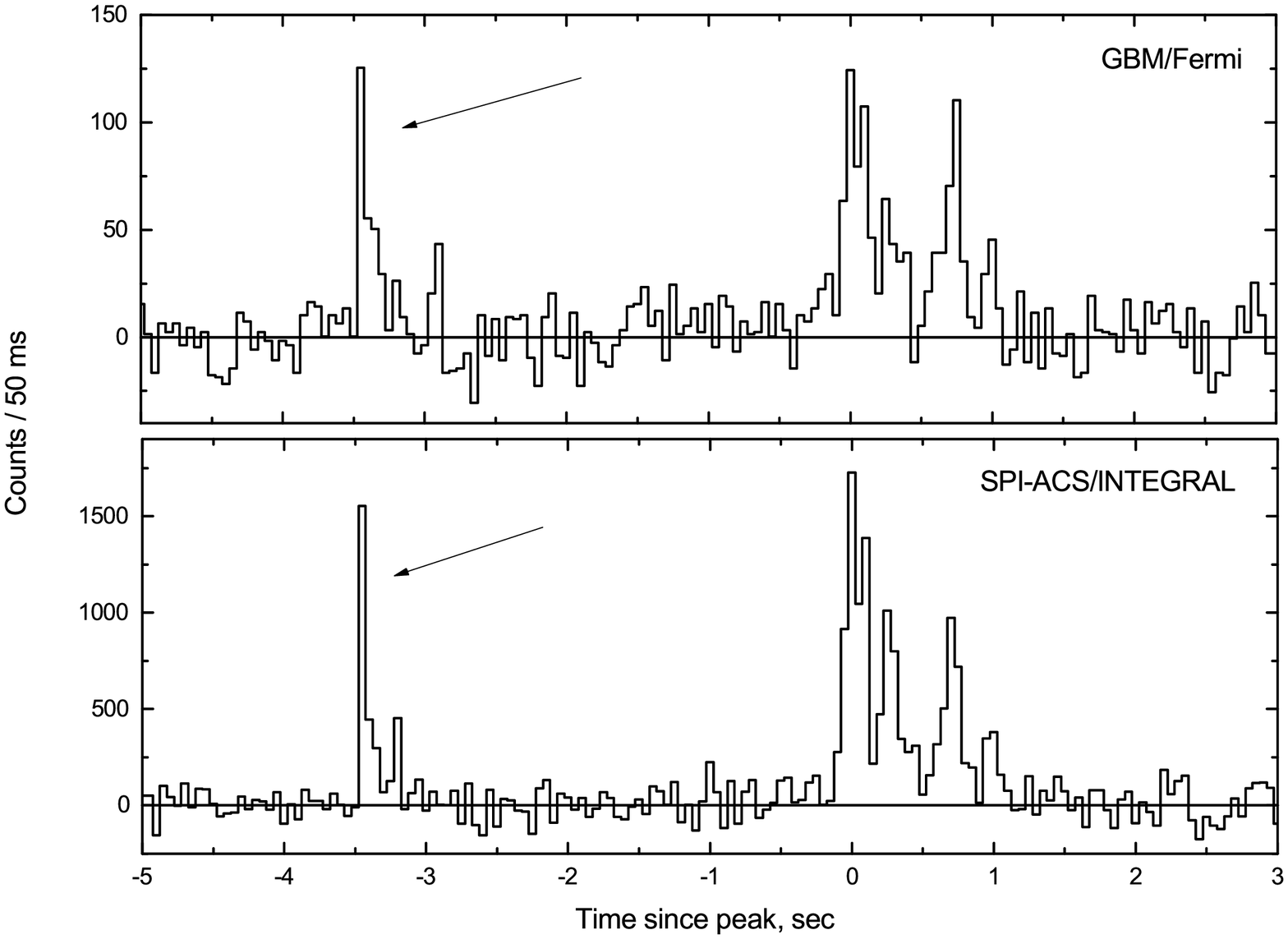}
\\ \textbf{Fig. 2.} \emph{Light curve of GRB 100717 from the SPI-ACS/INTEGRAL (bottom) and GBM/Fermi data in the energy range (0.1, 10) MeV (top). The GBM/Fermi light curve was constructed from the data of the NaI03, NaI07 – NaI11, and BGO01 detectors. The time relative to the peak in seconds is along the horizontal axis. The number of counts in 0.05 s is along the vertical axis. The arrow marks the precursor candidate.}
\end{figure}

\begin{figure}[h]
\epsfxsize=18cm \hspace{-0.7cm}\epsffile{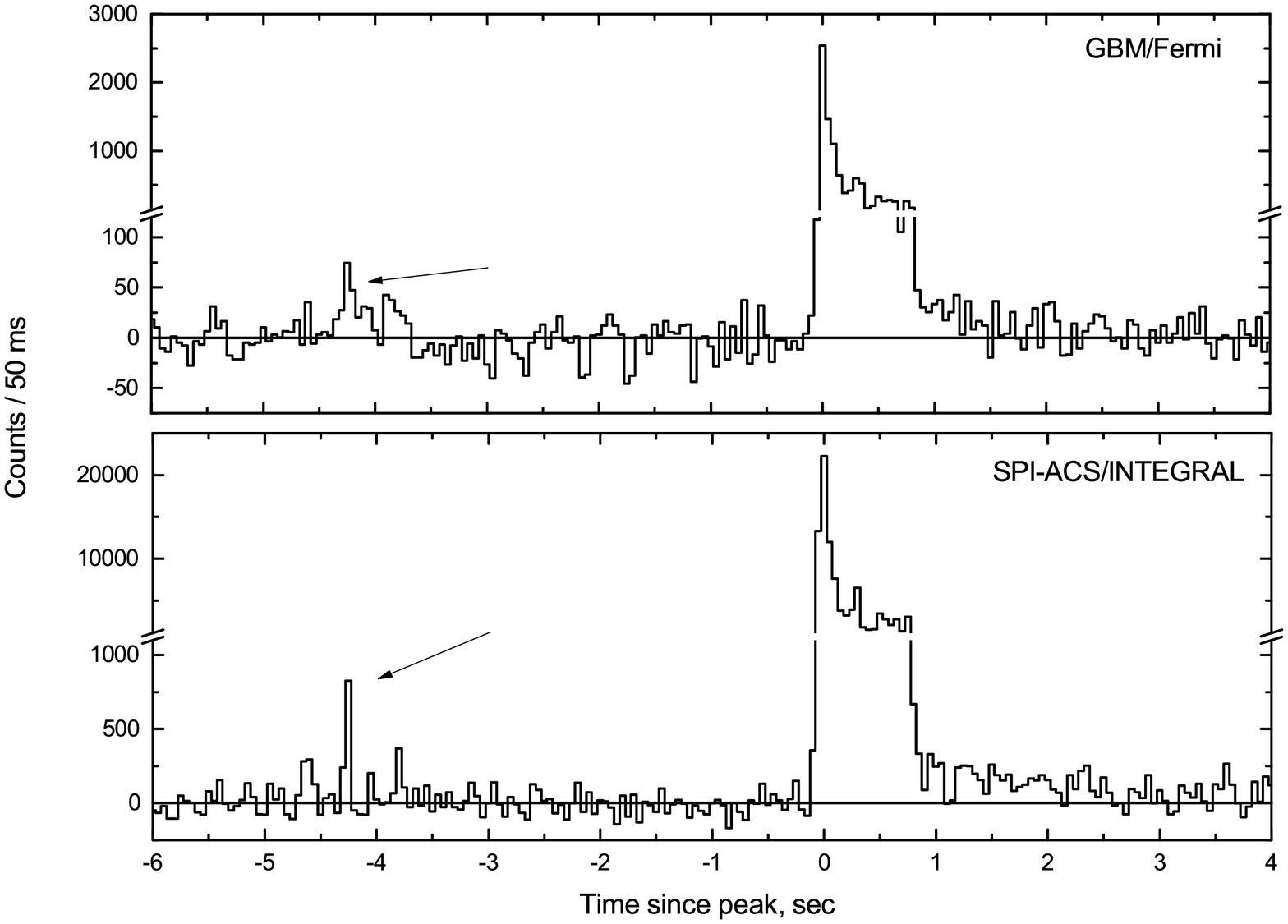}
\\ \textbf{Fig. 3.} \emph{Light curve of GRB 130310 from the SPI-ACS/INTEGRAL (bottom) and GBM/Fermi data in the energy range (0.1, 10) MeV (top). The GBM/Fermi light curve was constructed from the data of the NaI09 – NaI11, BGO00, and BGO01 detectors. The time relative to the peak in seconds is along the horizontal axis. The number of counts in 0.05 s is along the vertical axis. The arrow marks the precursor candidate.}
\end{figure}

\begin{figure}[h]
\epsfxsize=17cm \hspace{-0.3cm}\epsffile{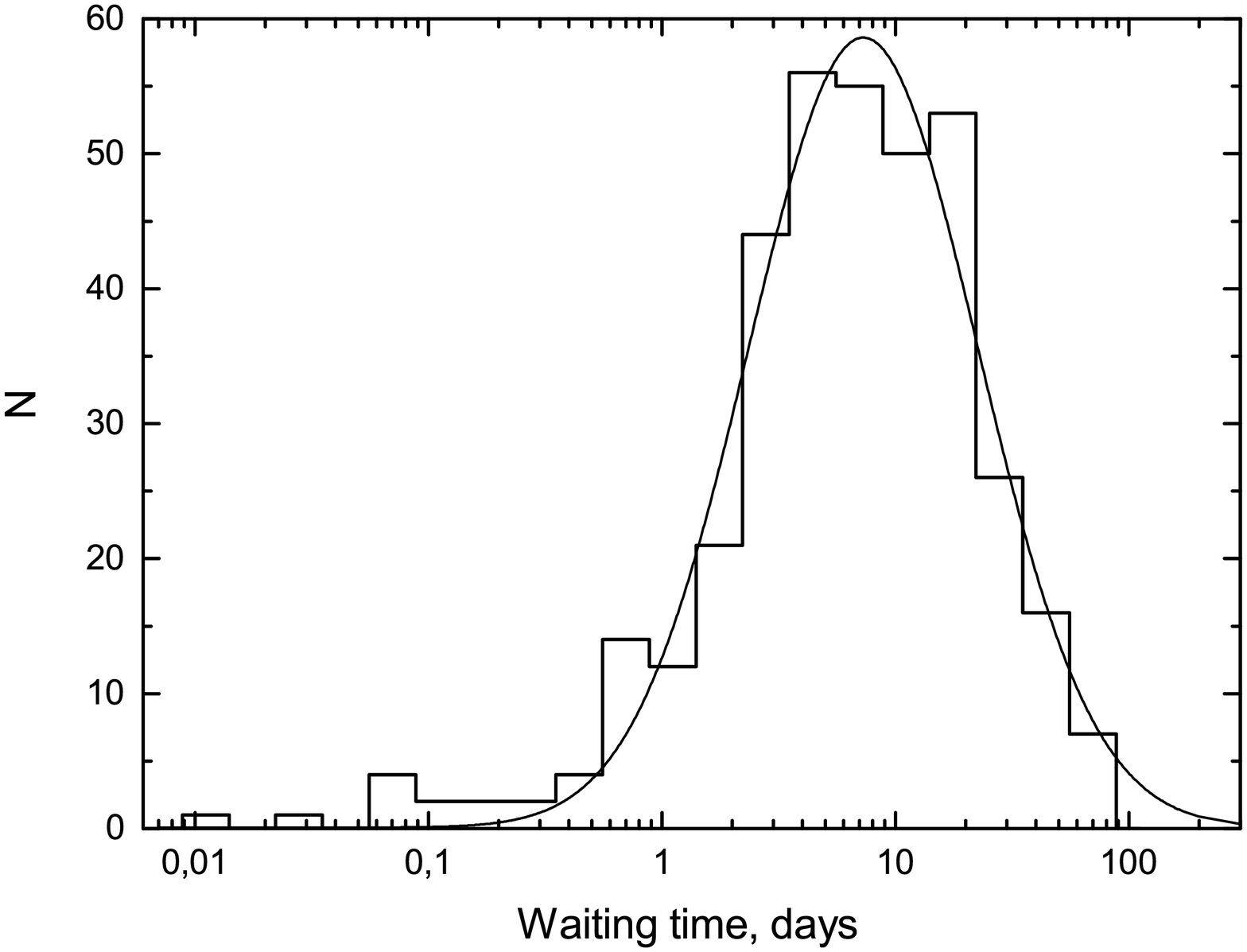}
\\ \textbf{Fig. 4.} \emph{Distribution of short GRBs in waiting time (the time interval between the successive detections of two short GRBs). The smooth curve indicates a log-normal fit. The waiting time in days is along the horizontal axis; the number of short GRBs is along the vertical axis.}
\end{figure}

\subsection*{Testing the Reliability of Precursor Candidates}

To estimate the probability that the detected precursor candidates are random background fluctuations, we can estimate the number of random spikes in the investigated time interval. In total, we analyzed $N=\frac{T\times N_{GRB}}{dT}=15570$ bins, where T is the time interval in which the precursor was searched for, dT is the time resolution of the investigated light curve, and $N_{GRB}$ is the number of analyzed GRBs. Assuming a normal distribution of counts in 15570 bins, the probability of a random background variation with a significance of six standard deviations is $1.5*10^{-5}$, i.e., it is fairly substantial for the precursor candidate for GRB 071030 with a significance of 6.3 $\sigma$ (the corresponding probability is $4.6*10^{-6}$). Therefore, the possibility that it is actually a background fluctuation must not be ruled out. The precursor candidates for GRB 100717 and GRB 130310 were detected at a higher significance level (12.8 and 10 $\sigma$, respectively); therefore, their association with the background fluctuations can be ruled out.

Let us estimate the probability that the precursor candidates are associated with other short GRBs. Let us construct the distribution of short GRBs in waiting time, i.e., the time interval between the successive detections of two short GRBs (Fig. 4). In several papers (McBreen et al. 1994; Pozanenko and Shatskiy 2010) the distribution constructed from the BATSE data was shown to be log-normal. Such a distribution is also traced for the short GRBs of the SPI-ACS/INTEGRAL experiment investigated here (the smooth curve in Fig. 4). It is important to note that the SPI-ACS experiment is switched off for the time of INTEGRAL passage through the radiation belts at the perigee of its orbit, on average, for 0.3 day with a 2.7-day interval. However, since the maximum of the distribution lies at 7.3 days (the position of the maximum of the log-normal model curve), this effect can be assumed to distort insignificantly the shape of the distribution (which is confirmed by good agreement between the model curve and observational data, Fig. 4). Thus, short GRBs are recorded in the SPI-ACS experiment, on average, every 7.3 days. We searched for precursor candidates in the time intervals (-50, -2) and (-5, -2) s for the light curves with a time resolution of 5 and 0.1 s, respectively. Assuming a Poissonian distribution, the probability of detecting at least one short GRB in these intervals is $7.6*10^{-5}$ and $4.8*10^{-6}$, respectively. Since we investigated 519 light curves, the probability of detecting a short GRB by chance in one of the intervals increases to $3.9*10^{-2}$ for the interval (-50, -2) s and to $2.5*10^{-3}$ for the interval (-5, -2) s. Thus, the probability that the detected precursor candidates are associated with other short GRBs is high, and this association must not be ruled out without additional studies of the candidates (for example, the localization of the sources of the precursor candidate and the main burst phase).

The count rate of events associated with the interaction of charged particles with the SPI-ACS detectors is quite high. Rau et al. (2005) showed that, on average, Np = 15 events per hour were recorded at a 4.5 $\sigma$ significance level. We searched for precursors in the light curves of 519 short bursts with a time resolution of 100 ms in the time interval (-5, -2) s relative to the peak in the light curve, which is $T = \frac{3\times519}{3600} = 0.43$ h. On average, $N=T \times Np \geq 6$ events are recorded in this time interval at a 4.5 $\sigma$ significance level. Such events are usually very short (the duration rarely exceeds 100 ms); the longest ones are simultaneously brightest and have a typical FRED shape (fast rise–exponential decay). The precursor candidate for GRB 071030 has a short duration and a low statistical significance. The possibility that this candidate is the interaction of charged particles with the detectors must not be ruled out, but this cannot be unequivocally asserted without any additional testing (for example, by comparison with the data of other experiments).

Thus, the precursor candidate for GRB 071030 is not reliable enough, because it can be associated with the interaction of charged particles with the detectors and with another short GRB. It may well be that it is a random background variation. We found the precursor candidates for GRB 100717 and GRB 130310 in the GBM/Fermi experiment (the precursor candidate for GRB 130310 is also mentioned in the corresponding GCN circular; Xiong and Chaplin 2013), which points to their reliability. The GBM/Fermi experiment allows the precursor candidate and the main phase of these GRBs to be analyzed in more detail. A detailed analysis will help answer the question of whether the candidates found actually differ in their properties from the main burst phase and may be deemed the sought-for precursors. Thus, consider the properties of GRB 100717 and GRB 130310 in more detail.
\\

\subsection*{GRB 100717}

\begin{figure}[h]
\epsfxsize=17cm \hspace{-0.3cm}\epsffile{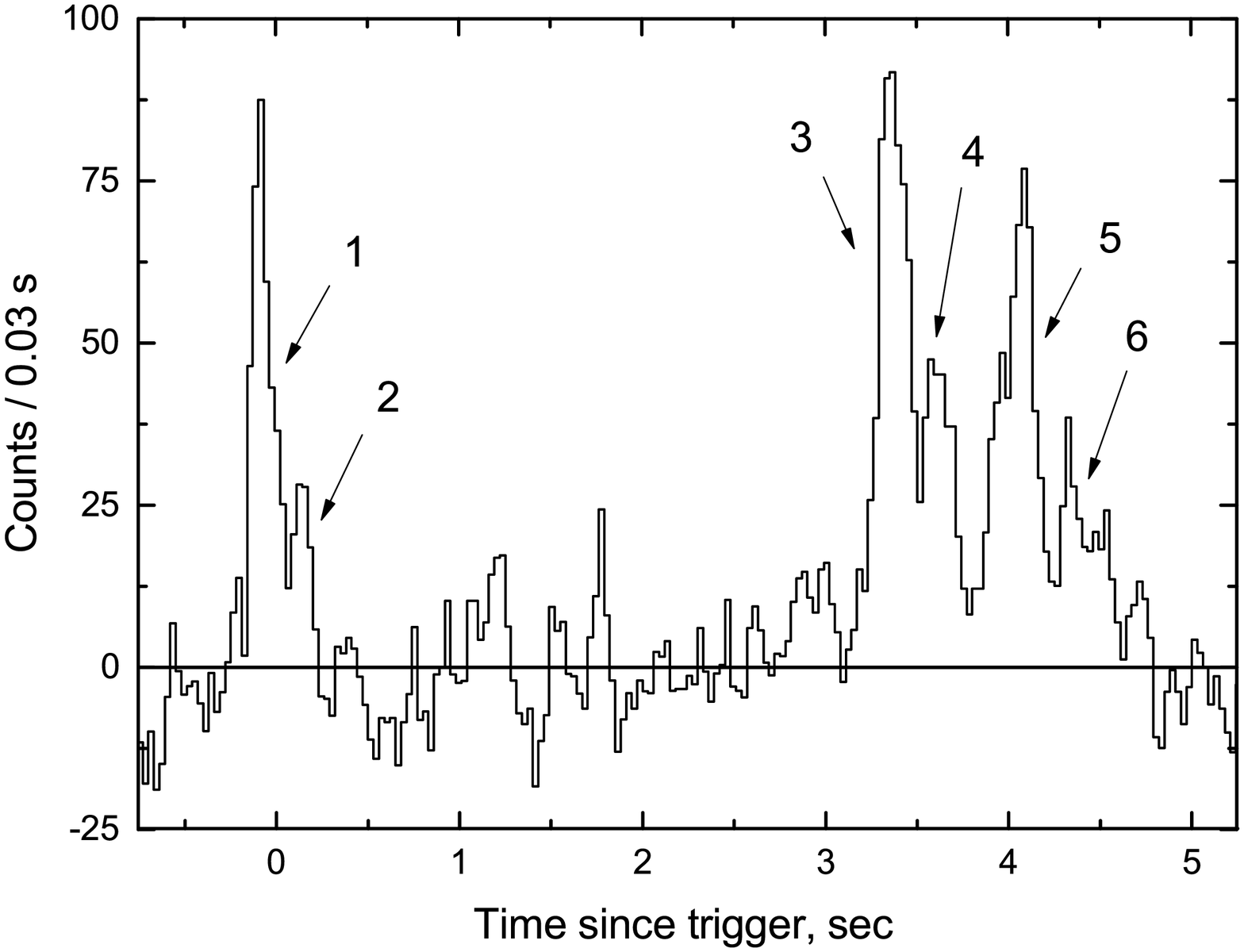}
\\ \textbf{Fig. 5.} \emph{Light curve of GRB 100717 from the data of the GBM/Fermi NaI03, NaI07 – Na I11, and BGO01 detectors in the energy range (8, 900) keV. The arrows indicate the individual pulses in the light curve (see also Table 2). The time relative to the GBN trigger in seconds is along the horizontal axis. The number of counts in 0.03 s is along the vertical axis.}
\end{figure}

We used the publicly accessible FTP archive (ftp://legacy.gsfc.nasa.gov/fermi/data/) as the source of the GBM and LAT of Fermi data used to analyze the GRBs.

Figure 5 presents the light curve of GRB 100717 in the energy range (8, 900) keV constructed from the data of the GBM/Fermi NaI03, NaI07 - NaI11, and BGO01 detectors with a time resolution of 0.03 s. The light curve of both the main GRB phase (the time interval (3, 5) s) and the precursor candidate (the time interval (-0.5, 0.5) s) consists of several pulses marked by the numbers. Interestingly, the automatic event search algorithm in the GBM/Fermi experiment triggered precisely at the precursor time, not at the main phase. The durations $T_{90}$, fluences, peak fluxes on a time scale of 30 ms, hardness ratios of the main phase and the precursor candidate as well as the individual pulses are given in Table 2. The hardness ratio was calculated as the ratio of fluences in the energy ranges (0.2, 0.9) MeV and (8, 200) keV expressed in raw counts, i.e., without allowance for the detector response matrix, and is given only for the possibility of comparing the values obtained for the precursor candidate and the main burst phase. For the bright events for which our spectral analysis was performed, the hardness ratio was also calculated using the response matrix and an optimal spectral model. The properties of the precursor candidate for GRB 100717 differ insignificantly from those of the main phase, especially if the properties of the individual pulses in the light curve are considered. This suggests that the detected precursor candidate is an early beginning of the main phase and that GRB 100717 is actually a long burst consisting of two groups of pulses separated by a “silence” episode with a total prompt phase duration $T_{90} = 4.6 \pm 0.1$ s.

For the light curves in the energy ranges (8, 200) and (200, 900) keV we performed a cross-correlation analysis similar to the technique of Band (1997). It showed that there is no statistically significant spectral lag between these light curves. Such a behavior is more typical for the class of short GRBs. However, in this case, it is probably not related to the nature of the phenomenon but is a methodological superposition effect that arises when a GRB with a complex multipeaked structure of the light curve is investigated (for more details, see Minaev et al. 2014).

\begin{table}[t]
\scriptsize
\vspace{6mm} \centering {{\bf Table 2.} \emph{The properties of the precursor candidates and the main GRB phase in the energy range (8, 900) keV based on GBM/Fermi data}}\label{meansp}

\vspace{5mm}\begin{tabular}{c|c|c|c|c|c|c} \hline\hline
GRB	   & Light curve        &Significance, & Т$_{90}$,      & Fluence,      &  Peak flux,        & Hardness       \\
       &  component $^{1}$  &$\sigma$&  s           & $10^{2}$ cnts       &  $10^{3}$ cnts/s $^{2}$       & ratio $^{3}$    \\\hline

090510 & precursor        & 13.2 & $0.05\pm0.02$  & $3.0\pm0.4$    & $12.8\pm1.3$ & $0.45\pm0.11$ \\
       & main phase       & 82.7 & $0.98\pm0.07$  & $79.2\pm1.9$   & $43.7\pm1.9$ & $0.46\pm0.02$ \\\hline
100717 & precursor        & 9.5& $0.30\pm0.05$  & $4.9\pm0.5$    & $2.9\pm0.4$ & $0.57\pm0.14$ \\
       & pulse 1         & - & $0.15\pm0.05$  & $3.8\pm0.4$    & $2.9\pm0.4$ & - \\
       & pulse 2         & - & $0.11\pm0.07$  & $1.1\pm0.3$    & $0.9\pm0.4$ & - \\
       & main phase      & 18.7& $1.4\pm0.2$    & $17.7\pm1.2$  & $3.1\pm0.4$ & $0.55\pm0.06$ \\
       & pulse 3         & - & $0.21\pm0.05$  & $6.0\pm0.5$    & $3.1\pm0.4$ & - \\
       & pulse 4         & - & $0.19\pm0.05$  & $2.9\pm0.5$    & $1.6\pm0.4$ & - \\
       & pulse 5         & - & $0.30\pm0.05$  & $5.8\pm0.6$    & $2.6\pm0.5$ & - \\
       & pulse 6         & - & $0.45\pm0.15$  & $3.0\pm0.6$    & $1.3\pm0.5$ & - \\\hline
130310 & precursor       & 7.1 & $0.9\pm0.2$    & $9.8\pm1.3$    & $2.0\pm0.4$ & $0.31\pm0.10$ \\
       & main phase     &134.1 & $2.7\pm0.5$    & $285.3\pm3.7$  & $49.3\pm0.7$ & $0.37\pm0.01$ \\

\hline
\multicolumn{7}{l}{}\\
\multicolumn{7}{l}{$^{1}$ The values for the individual pulses of the light curve are also specified for GRB 100717.}\\
\multicolumn{7}{l}{$^{2}$ The peak flux was obtained in the light curve with a time resolution of 15 ms for GRB 090510,}\\
\multicolumn{7}{l}{30 ms for GRB 100717, 150 ms for GRB 130310.}\\
\multicolumn{7}{l}{$^{3}$ The ratio of the fluences in the energy ranges (0.2, 0.9) MeV and (8, 200) keV expressed in raw counts.}\\

\end{tabular}
\end{table}

To construct and fit the energy spectra, we used the RMfit v4.3.2 software package specially developed to analyze the GBM and LAT/LLE data of the Fermi observatory (http://fermi.gsfc.nasa.gov/ssc/data/analysis/rmfit/). The method of spectral analysis is analogous to that proposed by Gruber et al. (2014), who also used the RMfit software package. Since the energy spectra are fitted in the RMfit software package with the maximum possible spectral resolution, the number of counts in each energy channel is small and, as a consequence, is not described by a normal distribution. Therefore, to fit the energy spectra and to choose an optimal spectral model, we used modified Cash statistics (CSTAT; see Cash 1979) instead of the $\chi^2$ test.

Based on the data of the GBM/Fermi NaI08, NaI11, and BGO00 detectors, we constructed the energy spectra of the main phase and the precursor candidate for GRB 100717, which were fitted by three spectral models: a simple power law (PL), a power law with an exponential cutoff (CPL), and a power law with a break (Band et al. 1993). The results of our spectral analysis are presented in Table 3. The energy spectrum of both the main phase and the precursor candidate is described unsatisfactorily by the simple power-law model. The optimal models of the energy spectrum for the main phase and the precursor candidate are, respectively, the power law with a break (Band) and the power law with an exponential cutoff (CPL). The Band model fits the energy spectrum of the precursor candidate bad due to its low statistical significance and, therefore, was not included in Table 3. Although the energy spectra are fitted by different spectral models, the parameters $E_{peak}$ and $\alpha$ coincide within the 1 $\sigma$ error limits, suggesting that the energy spectrum of the precursor candidate is analogous to that of the main phase. This is also confirmed by the hardness ratio calculated between the ranges (50, 300) and (10, 50) keV within the optimal model of the energy spectrum, $HR = 1.4 \pm 0.4$ for the main phase and $HR = 2.0 \pm 1.2$ for the precursor. Von Kienlin et al. (2014) constructed a duration - hardness diagram for GBM/Fermi GRBs. Although the hardness ratio in this paper was calculated by a slightly different method, the location of GRB 100717 on this two-dimensional diagram can be roughly estimated: it lies in the region of soft long bursts. The results of our spectral analysis for the main phase of GRB 100717 are consistent with those from Gruber et al. (2014).

Thus, GRB 100717 most likely belongs to the class of long bursts and, therefore, is not a sought-for event in this paper.

\subsection*{GRB 130310}

\begin{figure}[h]
\epsfxsize=17cm \hspace{-0.3cm}\epsffile{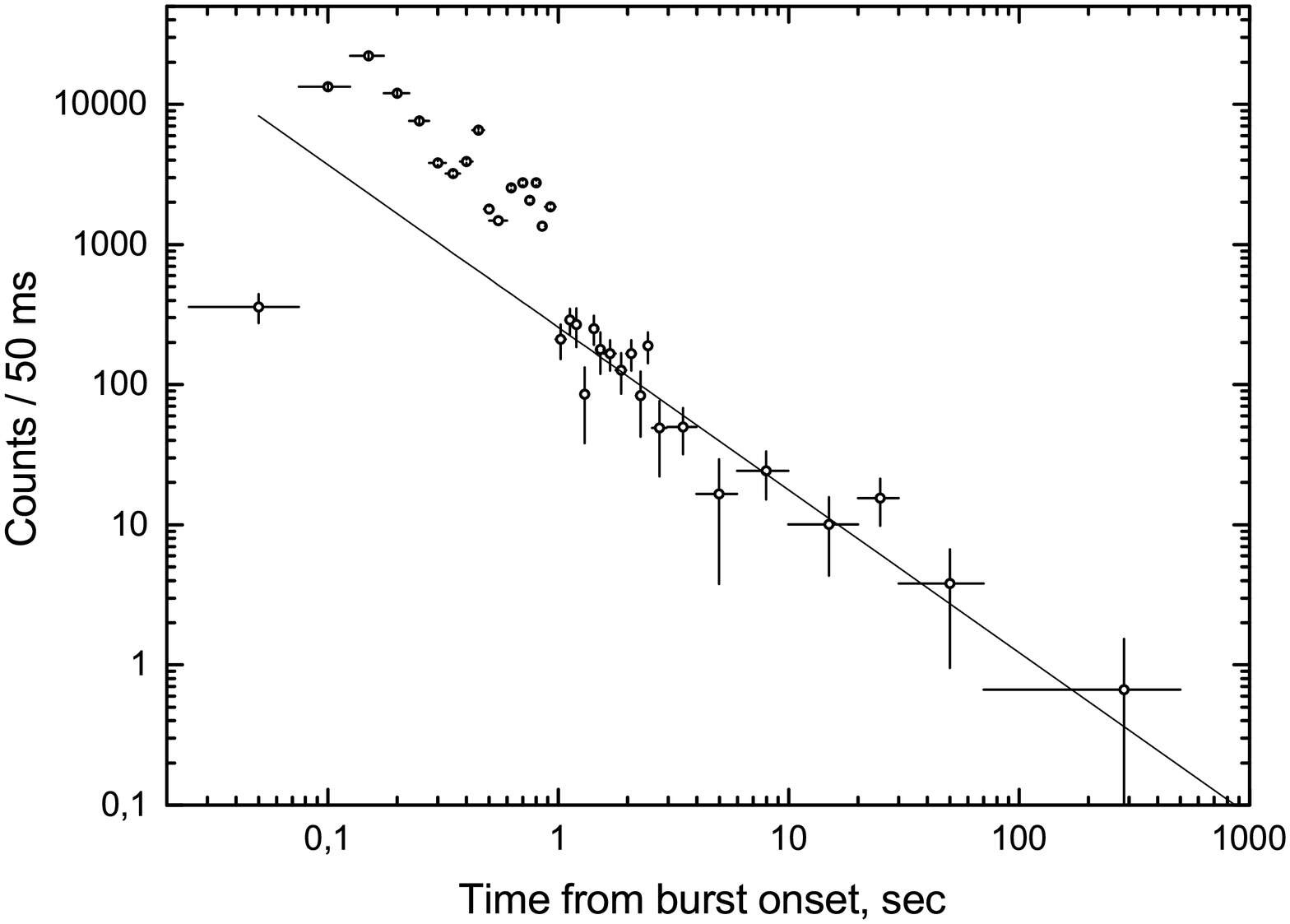}
\\ \textbf{Fig. 6.} \emph{Light curve of GRB 130310 from the SPI-ACS/INTEGRAL data. The straight line indicates the fit to the light curve in the interval (1, 500) s by a power law with an index $\alpha=-1.16\pm0.13$. The time relative to the burst onset in seconds is along the horizontal axis. The number of counts in 0.05 s is along the vertical axis.}
\end{figure}

The light curve of GRB 130310 differs significantly from that of GRB 100717. Both the fluence and the peak flux of the precursor candidate are much lower than those of the main burst phase (by a factor of 30  and 25, respectively; Table 2). At the same time, the duration of the precursor differs insignificantly from that of the main phase (0.9 and 2.7 s, respectively). The hardness ratio is the same within the 1 $\sigma$ error limits (Table 2).

\begin{table}[t]
\scriptsize 
\vspace{6mm} \centering {{\bf Table 3.} \emph{Results of the spectral analysis based on GBM and LAT/LLE of Fermi data}}\label{meansp}

\vspace{5mm}\begin{tabular}{c|c|c|c|c|c|c|c|c} \hline\hline
GRB	   &  Light curve 	& Spectral   & $\alpha$                 &  $\beta$                    & E$_{peak}$ $^{2}$,     &  $\gamma$                   & Fluence,                     & CSTAT/dof      \\
       &  component     &  model $^{1}$ &                          &                             &   MeV                  &                             & (10-1000) keV,             &                \\
       &        &                &                          &                             &                        &                             &  $10^{-6}$ erg cm$^{-2}$    &           \\\hline

090510 & main phase     & PL             &                         &                            &                          &   $-1.670^{+0.008}_{-0.003}$ & $1.82\pm0.03$              &  4409.3/849 \\
       &      & СPL            & $-0.90^{+0.01}_{-0.03}$ &                            & $5.82^{+1.10}_{-0.02}$   &                              & $4.08\pm0.06$              &  4238.5/848 \\
       &               & Band           & $-0.73^{+0.03}_{-0.02}$ & $-2.62^{+0.04}_{-0.04}$    & $2.83^{+0.17}_{-0.17}$   &                              & $4.52\pm0.09$              &  1070.2/847 \\
       &               & CPL+PL         & $-0.69^{+0.04}_{-0.03}$ &                            & $4.36^{+0.12}_{-0.12}$   & $-1.58^{+0.03}_{-0.02}$      & $4.28\pm0.07$              &  946.3/846 \\
       &               & Band+PL        & $-0.66^{+0.05}_{-0.04}$ & $-3.7^{+0.2}_{-0.3}$       & $3.9^{+0.2}_{-0.2}$      & $-1.60^{+0.03}_{-0.03}$      & $4.38\pm0.08$        &  928.3/845\\\cline{2-9}
       & precursor   & PL             &                         &                            &                          & $-1.30^{+0.05}_{-0.04}$      & $0.11\pm0.01$            &  746.4/818 \\
       &               & СPL            & $-0.6^{+0.3}_{-0.5}$    &                            & $0.8^{+2.2}_{-0.5}$      &                              & $0.16\pm0.02$            &  719.5/817 \\
       &               & Band           & $-0.5^{+0.4}_{-0.2}$    & $-2.0^{+0.3}_{-0.4}$       & $0.7^{+0.5}_{-0.3}$      &                              & $0.15\pm0.02$            &  715.1/816 \\
       &               & kT+PL          & $-1.4^{+0.1}_{-0.1}$    &                            & $0.12^{+0.03}_{-0.02}$   &                              & $0.15\pm0.02$      &  724.0/816\\\cline{2-9}
       & extended emis.    & PL             &                         &                            &                          & $-1.55^{+0.01}_{-0.01}$      & $0.85\pm0.07$              &  1148.4/609 \\\hline
100717 & main phase      & PL             &                         &                            &                          &  $-1.28^{+0.02}_{-0.03}$     & $1.21\pm0.06$              &  492.8/349 \\
       &     & СPL            & $-0.96^{+0.08}_{-0.08}$ &                            & $4.1^{+1.8}_{-1.0}$      &                              & $1.61\pm0.09$              &  458.3/348 \\
       &               & Band           & $-0.71^{+0.15}_{-0.14}$ & $-1.71^{+0.10}_{-0.15}$    & $1.2^{+0.7}_{-0.8}$      &                              & $1.66\pm0.11$        &  446.1/347\\\cline{2-9}
       & precursor   & PL             &                         &                            &                          &  $-1.35^{+0.06}_{-0.05}$     & $0.17\pm0.02$            &  421.8/349 \\
       &               & СPL            & $-0.4^{+0.4}_{-0.3}$    &                            & $0.7^{+0.6}_{-0.4}$      &                              & $0.26\pm0.03$            &  401.8/348\\\hline
130310 & main phase      & PL             &                         &                            &                          & $-1.448^{+0.005}_{-0.003}$   & $7.70\pm0.07$             &  3508.5/592 \\
       &           & СPL            & $-1.102^{+0.010}_{-0.009}$ &                         & $3.99^{+0.18}_{-0.19}$   &                              & $11.31\pm0.12$             &  852.2/591 \\
       &               & Band           & $-1.033^{+0.015}_{-0.015}$ & $-2.52^{+0.08}_{-0.09}$ & $2.23^{+0.21}_{-0.20}$   &                              & $11.79\pm0.14$             &  754.7/590 \\
       &               & CPL+PL         & $-1.052^{+0.025}_{-0.016}$ &                         & $2.94^{+0.19}_{-0.21}$   & $-1.43^{+0.08}_{-0.07}$      & $11.60\pm0.14$       &  790.8/589 \\\cline{2-9}
       & main peak       & PL             &                         &                            &                          &  $-1.322^{+0.008}_{-0.008}$  & $2.27\pm0.04$            &  847.2/578 \\
       &      & СPL            & $-1.184^{+0.014}_{-0.014}$ &                         & $11.8^{+1.4}_{-1.2}$     &                              & $2.69\pm0.06$              &  693.9/577 \\
       &               & Band           & $-1.180^{+0.012}_{-0.014}$ & $-3.4^{+0.7}_{-1.3}$    & $12.4^{+1.9}_{-1.6}$     &                              & $2.71\pm0.06$              &  695.5/576 \\
       &               & CPL+PL         & $-0.67^{+0.16}_{-0.17}$ &                            & $6.1^{+1.5}_{-1.0}$      &  $-1.56^{+0.05}_{-0.14}$     & $2.61\pm0.06$              &  639.9/575 \\
       &               & Band+PL        & $-0.71^{+0.16}_{-0.13}$ & $-2.5^{+0.2}_{-0.4}$       & $5.5^{+1.0}_{-0.8}$      &  $-1.68^{+0.11}_{-0.15}$     & $2.65\pm0.07$        &  635.7/574\\\cline{2-9}
       & extended emis.   & PL             &                         &                            &                          &    $-1.75^{+0.03}_{-0.03}$   & $0.99\pm0.06$              &  840.6/592 \\\hline

\multicolumn{9}{l}{}\\
\multicolumn{9}{l}{$^{1}$ PL is a power law, CPL is a power law with an exponential cutoff, Band is a power law with a break (Band et al. 1993),}\\
\multicolumn{9}{l}{kT is a thermal model.}\\
\multicolumn{9}{l}{$^{2}$ For the kT spectral model the E$_{peak}$ column gives the parameter kT.}\\

\end{tabular}
\end{table}

A more in-depth analysis of the light curve based on SPI-ACS/INTEGRAL data revealed a third component, an extended emission, which is a weak “tail” with a duration of 500 s (Fig. 6). The light curve of the extended emission was fitted in the interval (1, 500) s relative to the burst onset by a power law with an index $\alpha=-1.16\pm0.13$. Thus, the light curve of GRB 130310 consists of three components: the precursor candidate, the main phase with a duration of $\simeq$ 1 s, and the extended emission with a duration of $\simeq $ 500 s. Due to the extended emission the duration parameter $T_{90}$ is $2.7\pm0.5$ s, which formally characterizes GRB 130310 as a long burst if the separation criterion $T_{90}$ = 2 s is used. However, if the duration of the main phase ($\simeq$ 1 s) is taken as the burst duration, then the burst more likely belongs to the class of short ones. There is no spectral lag between the light curves in the ranges (8, 200) and (200, 900) keV, which, as in the case of GRB 100717 may be associated with the complex structure of its light curve consisting of several strongly overlapping pulses (Fig. 6). 

To investigate the energy spectrum, we jointly used the data of the GBM NaI09 - NaI11, BGO00, and BGO01 detectors and those from the Fermi LAT/LLE (LAT Low Energy) experiment. As a result, we constructed the energy spectrum of the main phase (the time interval (0.0, 1.0) s relative to the peak in the light curve) in the wide energy range 10 keV–100 MeV (Fig. 7a, Table 3), which was fitted by four models: a simple power law (PL), a power law with an exponential cutoff (CPL), a power law with a break (Band), and the sum of the simple power law and the power law with an exponential cutoff (PL + CPL). The optimal model is the power law with a break (Band). The peak in the energy spectrum $\nu F_{\nu}$ is placed at $E_{peak} = 2.23^{+0.21}_{-0.20}$ MeV, the fluence in the energy range (10, 1000) keV is $F = (11.79\pm0.14)\ast 10^{-6}$ erg cm$^{-2}$. The hardness ratio between the ranges (50, 300) and (10, 50) keV within the optimal spectral model is $HR = 1.01 \pm 0.02$. On the duration - hardness diagram constructed for GBM/Fermi GRBs (Fig. 6 in von Kienlin et al. (2014)) GRB 130310 is intermediate between the soft long and hard short bursts, which complicates its classification. The results of our spectral analysis for the main phase of GRB 130310 are consistent with those from Gruber et al. (2014).

There is an interesting feature in the light curve for the main phase of GRB 130310: it begins with a very short and bright pulse $\simeq$ 0.1 s in duration (Fig. 3). Therefore, we constructed the separate energy spectrum for it from the data of the GBM/Fermi NaI09 – NaI11, BGO00, and BGO01 detectors presented in Fig. 7b. The optimal spectral model is the sum of a power law with an index $\gamma = -1.68^{+0.11}_{-0.15}$ and a power law with a break (Band et al. 1993) with a very high value of the parameter $E_{peak} = 5.5^{+1.0}_{-0.8}$ MeV (Table 3). 

\epsfxsize=10cm \hspace{2cm}\epsffile{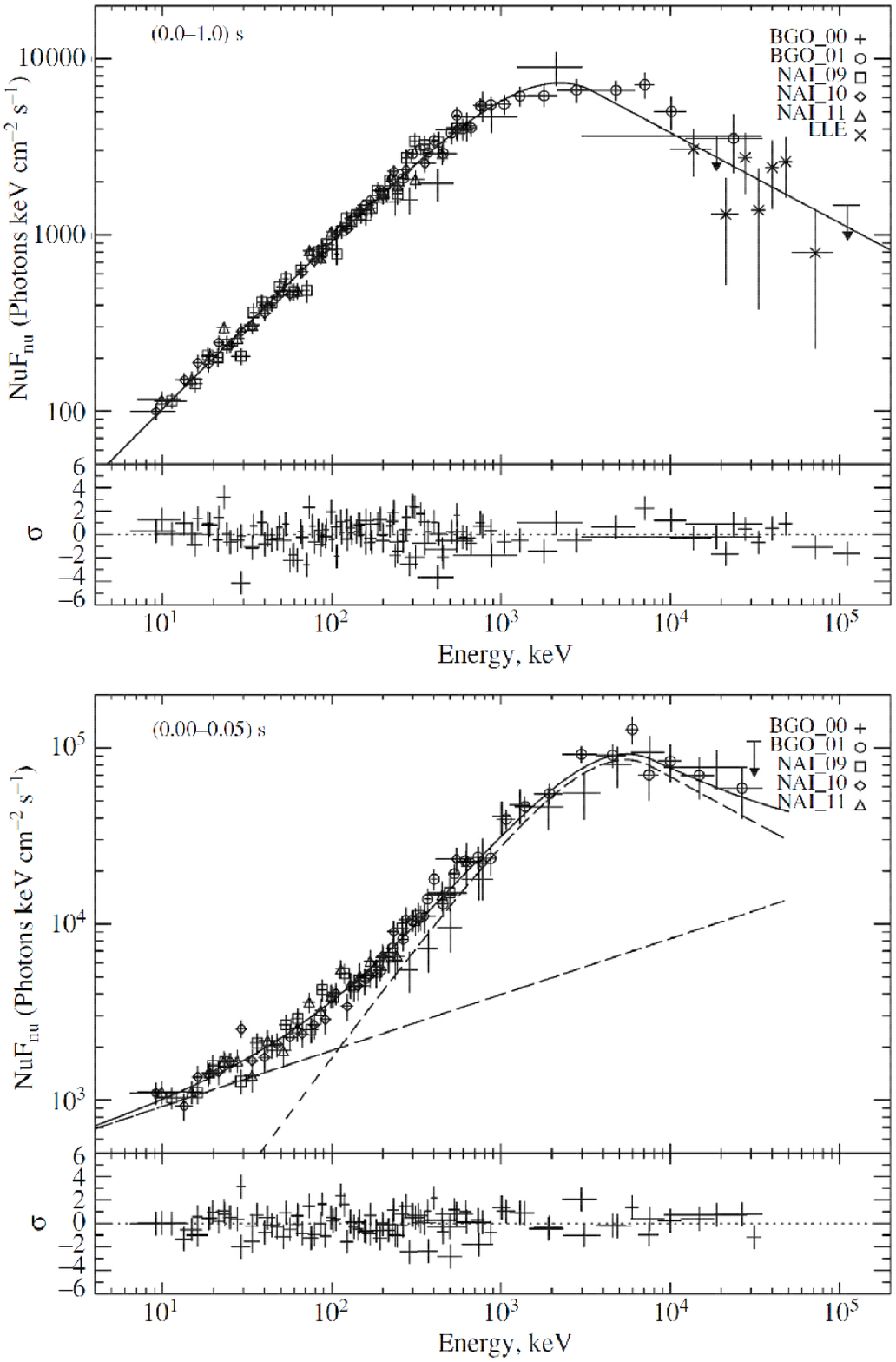}
\\ \textbf{Fig. 7.} \emph{Energy spectrum $\nu F_{\nu}$ of GRB 130310. (a) The energy spectrum constructed from the data of the GBM NaI09 - NaI11, BGO00, and BGO01 detectors and the Fermi LAT/LLE data in the time interval (0.0, 1.0) s relative to the peak in the light curve covering the main GRB phase. The smooth curve indicates the fit to the spectrum by a power law with a break (Band et al. 1993). The lower panel shows the deviation of the model from the experimental data expressed in standard deviations. (b) The energy spectrum of the main peak constructed from the data of the GBM NaI09 - NaI11, BGO00, and BGO01 detectors spanning the time interval (0.00, 0.05) s relative to the peak; the smooth curve indicates the fit by the sum of a power law and a power law with a break; the contribution of the individual components to the combined spectrum is also shown. The lower panel shows the deviation of the spectral model from the experimental data expressed in standard deviations.}

Such a complex shape of the energy spectrum was observed only in a few cases (see, e.g., Ackermann et al. 2010; Svinkin et al. 2016; and the next section in this paper).

We also constructed the energy spectrum of the extended emission (the time interval (1, 4) s relative to the peak in the light curve) from the data of the NaI09 - NaI11, BGO00, and BGO01 detectors and the Fermi LAT(LLE) data, which was fitted by a simple power law with an index $\gamma = -1.75 \pm 0.03$ (Table. 3). The introduction of additional free parameters by using more complex spectral models (CPL, Band, etc.) does not improve the quality of the fit due to the low statistical significance of the light-curve component being investigated. Interestingly, the derived power law index coincides, within the 1 $\sigma$ error limits, with that of the additional component in the spectrum of the main peak. This suggests that the extended emission is an additional component of the light curve that is also present in the main burst phase and probably has a different emission mechanism compared to the main phase. However, a detailed study of this component is beyond the scope of this paper and is the subject of future studies.

Since the statistical significance of the precursor candidate is low, its energy spectrum cannot be constructed and compared with the spectrum of the main phase. We can only compare the hardness ratios calculated in raw counts; they coincide within the 1 $\sigma$ error limits. This may suggest that the precursor candidate is the beginning of the main GRB phase.

\begin{figure}[h]
\epsfxsize=17cm \hspace{-0.3cm}\epsffile{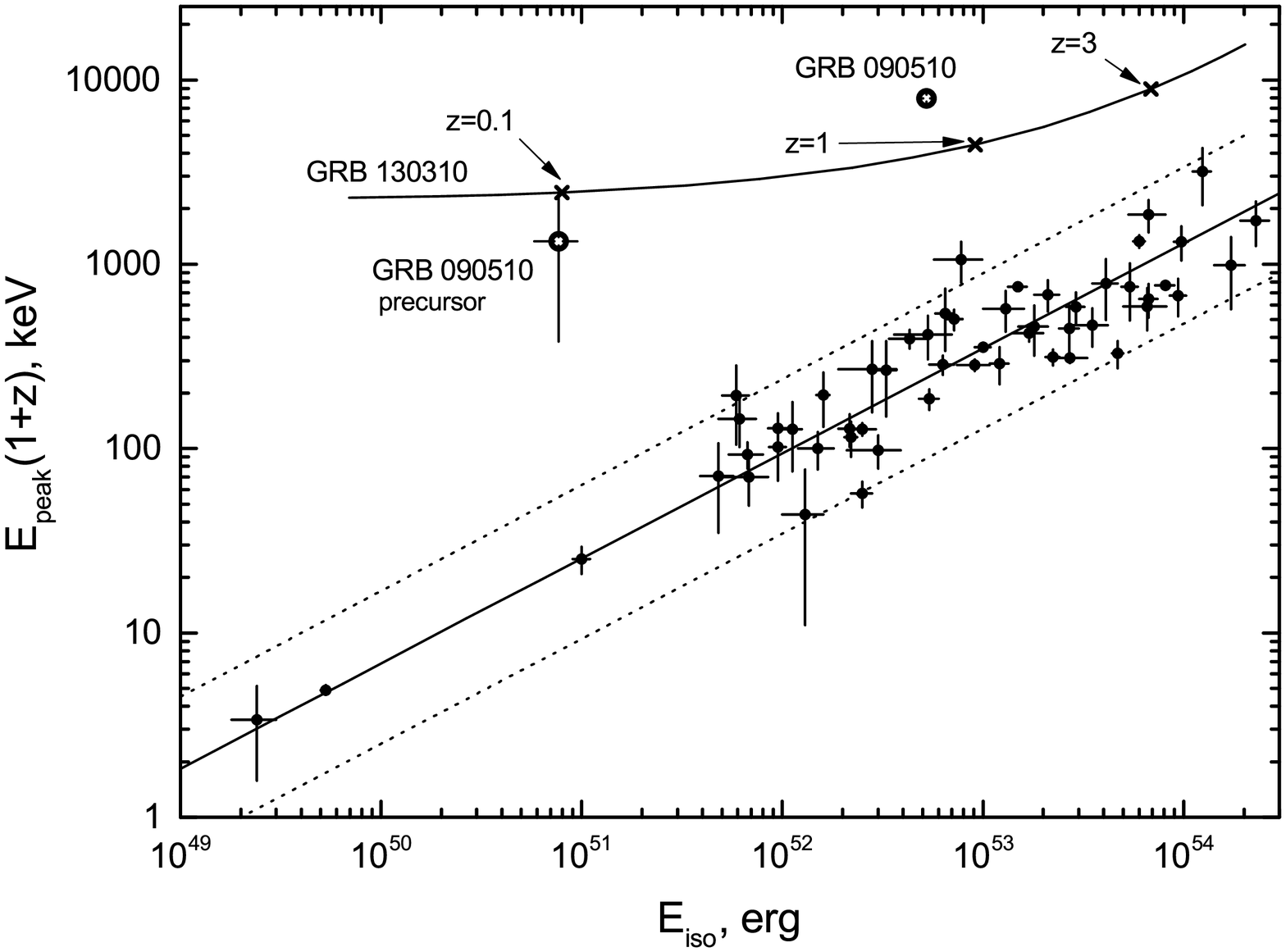}
\\ \textbf{Fig. 8.} \emph{Amati diagram - equivalent isotropic energy emitted in the gamma-ray range $E_{iso}$ versus parameter $E_{peak}(1 + z)$ in the source reference frame (Amati 2010). The solid straight line indicates a power law fit to the dependence; the dashed lines bound the 2 $\sigma$ correlation region. The trajectory of GRB 130310 is plotted as a function of the presumed redshift z. The filled circles indicate the data points for the main phase and the precursor candidate of GRB 090510 (the 1 $\sigma$ errors in the parameters $E_{iso}$ and $E_{peak}(1 + z)$ for the main phase of GRB 090510 are smaller than the circle size).}
\end{figure}

One of the interesting phenomenological relations for GRBs is Amati diagram, i.e., the dependence of the equivalent isotropic energy $E_{iso}$ emitted in the gamma-ray range (1, 10000) keV on parameter $E_{peak}(1 + z)$ in the source reference frame (Amati 2010). Long bursts obey well this law, while short bursts usually lie on the diagram above the main correlation region of long bursts (at the same value of $E_{iso}$ for short bursts $E_{peak}(1 + z)$ is considerably higher). Thus, Amati diagram can also be used for the classification of bursts. Since the redshift z of GRB 130310 is unknown, the trajectory of GRB 130310 on the diagram as a function of z (Fig. 8; see also Minaev et al. 2012) can be constructed using the fluence and $E_{peak}(1 + z)$ estimates. It follows from Fig. 8 that the trajectory does not cross the correlation region at any z and lies above it, which may suggest that GRB 130310 belongs to the class of short bursts. Since the burst does not fall into the $E_{peak}(1 + z) - E_{iso}$ correlation region at any z, the redshift and $E_{iso}$ cannot be estimated.

Thus, a detailed study of GRB 130310 revealed many peculiarities of both the light curve and the energy spectrum. However, it cannot be unambiguously said whether GRB 130310 is a short burst. Whether the precursor candidate has a different nature/emission mechanism compared to the main GRB phase or it is the beginning of the main phase is not clear either.

\subsection*{GRB 090510}

Let us consider one more short gamma-ray burst, GRB 090510. It was detected by most of the then operating space gamma-ray observatories, including INTEGRAL, Fermi, Swift, and KONUS (see Table 1). In Troja et al. (2010) it figures as a short GRB with a precursor. Moreover, Troja et al. (2010) found another precursor candidate in the BAT/Swift data 10 s before the beginning of the main phase, but it was not confirmed by the GBM/Fermi and SPI-ACS/INTEGRAL data. It may well be that it is a background fluctuation. However, in the SPI-ACS experiment we found the main precursor candidate at 0.5 s before the beginning of the main phase at a 4.3 $\sigma$ significance level (Fig. 9). It was not initially considered as a precursor candidate, because (1) its statistical significance was less than 6 $\sigma$ and (2) the time interval between it and the beginning of the main phase was less than 2 s.

\begin{figure}[h]
\epsfxsize=17cm \hspace{-0.3cm}\epsffile{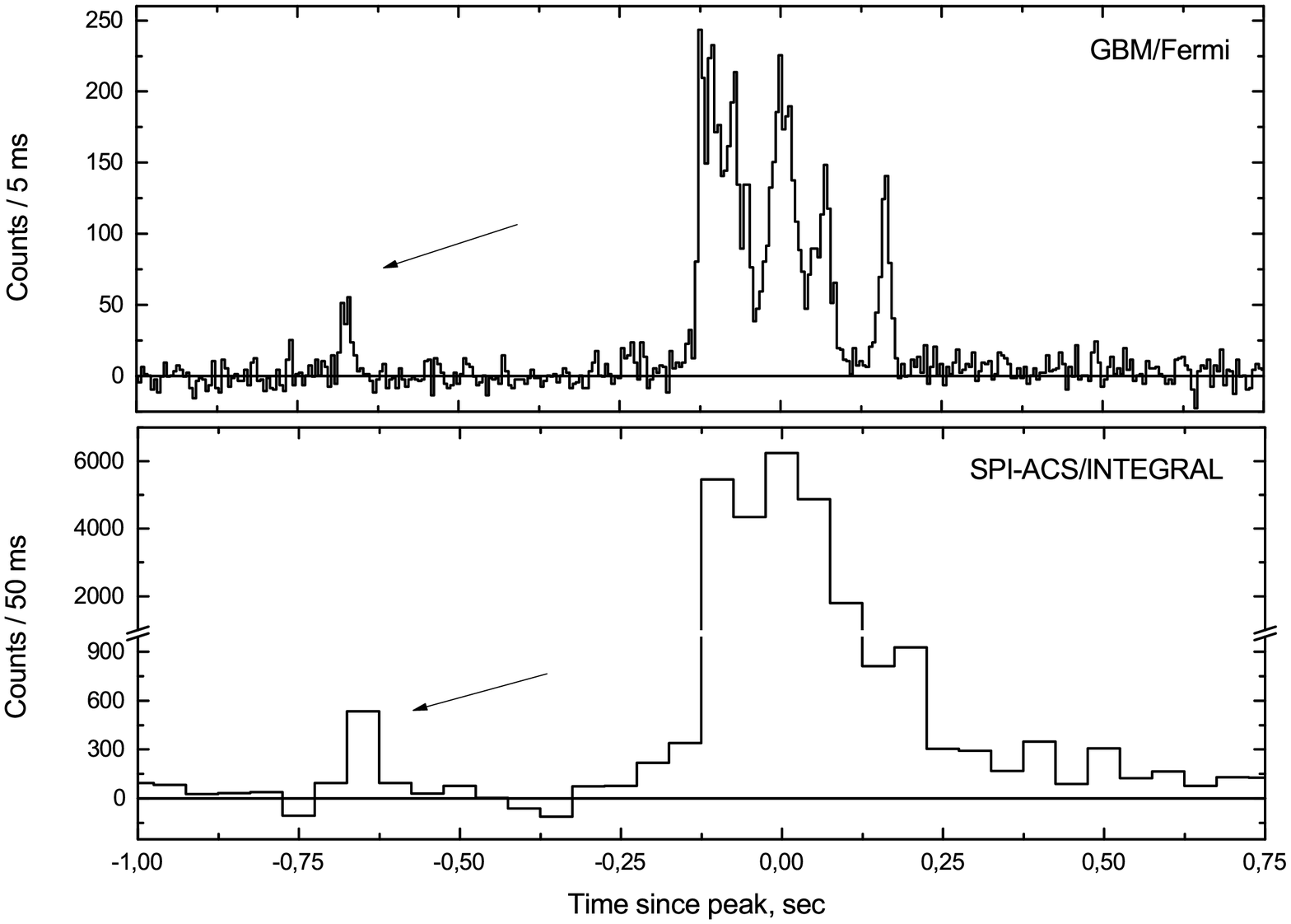}
\\ \textbf{Fig. 9.} \emph{Light curve of GRB 090510 from the SPI-ACS/INTEGRAL (bottom) and GBM/Fermi data in the energy range (0.1, 10) MeV (top). The GBM/Fermi light curve was constructed from the data of the NaI03, NaI06 - NaI09, BGO00, and BGO01 detectors. The time relative to the peak in seconds is along the horizontal axis. The number of counts in 0.05 (bottom) and 0.005 s (top) is along the vertical axis. The arrow marks the precursor candidate.}
\end{figure}

The low statistical significance of the precursor candidate in the SPI-ACS experiment can be partly associated with the low SPI-ACS efficiency toward the burst source: the angle between the SPI-ACS axis and the direction to the burst source is $\theta$ = 140.8$^\circ$. A detailed study of the dependence of the SPI-ACS detector efficiency on the direction to the GRB source is contained in Vigano and Mereghetti (2009), who showed the SPI-ACS efficiency for events with $\theta$ > 120$^\circ$ to be approximately half that for events with $\theta \sim 90 ^\circ$.

Consider the properties of the precursor candidate and the main phase of GRB 090510 in more detail. 

The light curve of GRB 090510 is very similar in shape to that of GRB 100717, the main phase consists of more than four overlapping pulses with a total duration $T_{90} = 0.98 \pm 0.07$ s, the duration of the precursor candidate $T_{90} = 0.05 \pm 0.02$ s coincides with that of the last pulse of the main phase (t = 0.15 s in Fig. 9). It was previously shown for GRB 100717 that the properties of the precursor candidates are similar to those of the individual pulses of the main phase. It may well be that the precursor candidate for GRB 090510 can also be an individual pulse of the main GRB phase.

The statistical significance of the precursor candidate in the GBM/Fermi data is sufficient to perform a spectral analysis. The energy spectrum of the precursor was constructed from the data of the NaI03, NaI06 - NaI09, BGO00, and BGO01 detectors and was fitted by four models: a simple power law (PL), a power law with an exponential cutoff (CPL), a power law with a break (Band), and the sum of a simple power law and a blackbody model (kT + PL). The optimal model is the power law with a break (Band); the kT + PL model with the same number of free parameters as the Band model fits the observed spectrum more poorly (Table 3). The parameters of the derived Band spectral model are typical for the class of short bursts (see, e.g., Gruber et al. 2014). The hardness ratio of the precursor candidate calculated using Band model is $HR = 1.9 \pm 0.7$. On the hardness – duration diagram constructed for GBM/Fermi GRBs (Fig. 6 in von Kienlin et al. (2014)) the precursor candidate for GRB 090510 is located in the region of very short hard bursts, i.e., occupies a position quite typical for the class of short GRBs. Thus, the spectral–temporal properties of the precursor candidate for GRB 090510
do not differ from those of the main phase of short GRBs, which may suggest that this component of the light curve is one of the pulses of the main phase of GRB 090510.

To construct the energy spectrum of the main phase, we used the data of the GBM NaI03, NaI06 – NaI09, BGO00, and BGO01 detectors and the LAT(LLE)/Fermi data. The energy spectrum of the main phase spanning the range 10 keV - 3 GeV is presented in Fig. 10. The energy spectrum of GRB 090510 is similar in shape to that of the main peak of GRB 130310 (Fig. 7b). The optimal spectral model is the sum of a power law with an index $\gamma = -1.60 \pm 0.03$ and a power law with a break (Band et al. 1993) with $E_{peak} = 3.9 \pm 0.2$ MeV (Table 3). The hardness ratio of the main phase of GRB 090510 calculated within the optimal Band + PL spectral model is $HR = 1.4 \pm 0.1$. On the hardness–duration diagram (Fig. 6 in von Kienlin et al. (2014)) the main phase of GRB 090510 is located in the region of short hard bursts.

\begin{figure}[h]
\epsfxsize=14cm \hspace{0.6cm}\epsffile{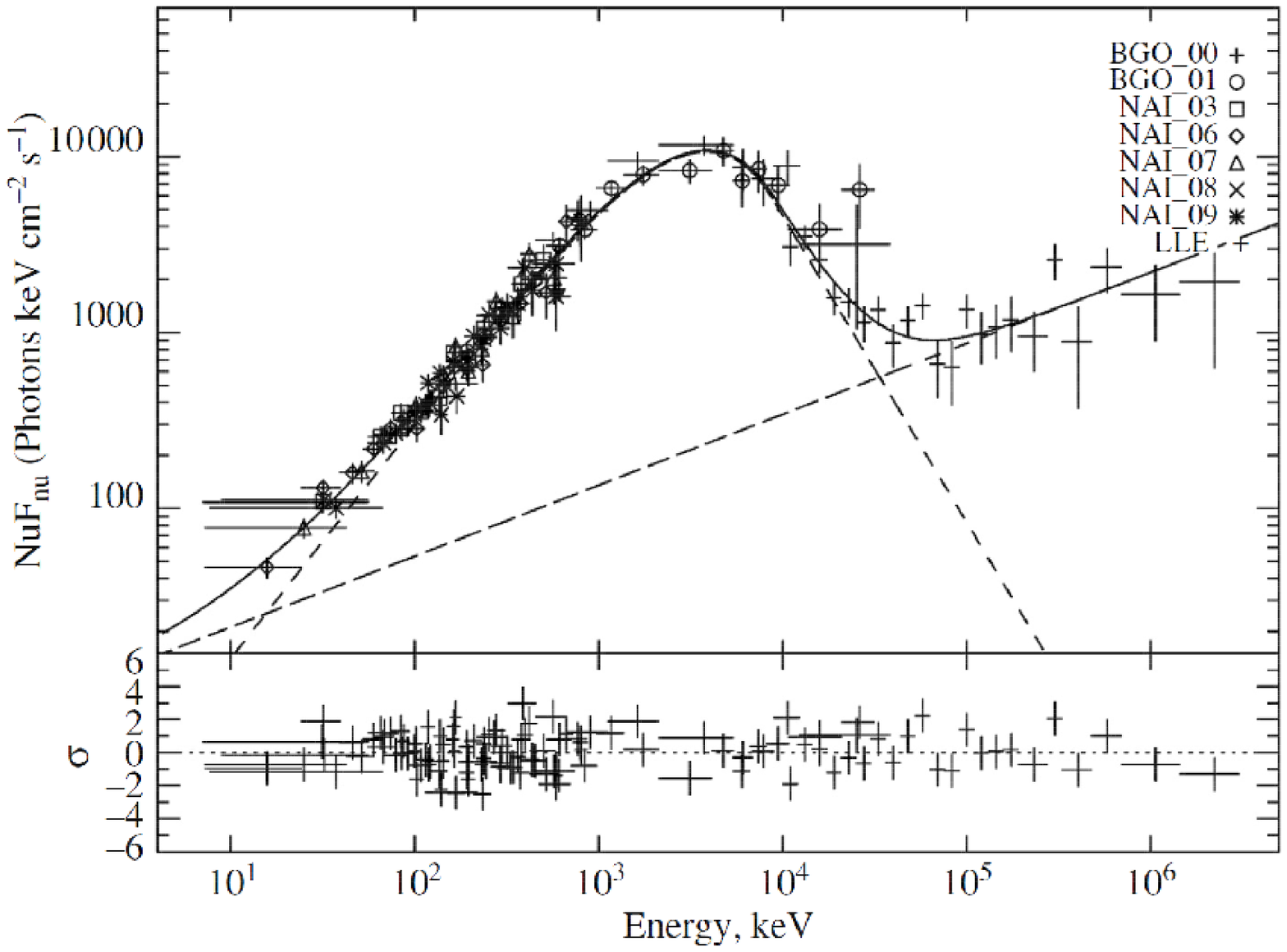}
\\ \textbf{Fig. 10.} \emph{Energy spectrum $\nu F_{\nu}$ of GRB 090510 constructed from the data of the GBM NaI03, NaI06 – NaI09, BGO00, and BGO01 detectors and the Fermi LAT(LLE) data covering the main GRB phase. The smooth curve indicates the fit by the sum of a power law and a power law with a break (Band et al. 1993); the contribution of the individual components to the combined spectrum is also shown. The lower panel shows the deviation of the spectral model from the experimental data expressed in standard deviations.}
\end{figure}

An additional power law component dominates in the energy spectrum for the main phase of GRB 090510 at energies below 10 keV and above 100 MeV. This component is usually associated with the high-energy emission (E > 1 GeV) from GRBs (see, e.g., Ackermann et al. 2010), which is characterized by a negative spectral lag compared to the sub-MeV emission, a longer duration, and a power law like flux decrease with time. This, in turn, suggests that the extended emission of GRB 130310 detected in the MeV energy range in the SPI-ACS experiment up to 500 s after the main burst phase is the most probably this component. An extended emission from GRB 090510 with a duration of $\simeq$ 10 s was detected in the GBM/Fermi data (Ackermann et al. 2010), although it was not identified there as a separate component of the light curve. In this paper we constructed its energy spectrum from the data of the GBM NaI06 - NaI09, and BGO01 detectors and the LAT(LLE)/Fermi data. The spectrum was fitted by a simple power law with an index $\gamma = -1.56 \pm 0.01$ (Table 3). Due to the low statistical significance of the extended emission in the GBM data (in the LAT(LLE) data the extended emission is considerably more intense), fitting the spectrum by more complex spectral models gives ambiguous results. Interestingly, the derived index of the power law spectral model coincides, within the error limits, with that of the additional component observed in the main burst phase. This may suggest that the extended emission from GRB 090510 (and GRB 130310) visible in sub-MeV energy range is associated with the high-energy (E > 1 GeV) emission component that is also observed in long bursts whose nature has not yet been clarified. In other words, the extended emission from GRB 090510 and
GRB 130310 is probably different in nature compared to the “classical” extended emission from short GRBs (see, e.g., Gehrels et al. 2006; Minaev et al. 2010a; Metzger et al. 2008; Barkov and Pozanenko 2011). This is also suggested by a different shape of the light curve for the extended emission: a power law monotonic flux decrease with time with a power law index close to unity is typical for GRB 090510 and GRB 130310 (Fig. 6). The classical extended emission observed in the light curves of some short bursts is characterized by a complex nonmonotonic structure (see, e.g., Gehrels et al. 2006).

The redshift of GRB 090510 is z = 0.903 (McBreen et al. 2010). Using a standard cosmological model ($[\Omega_{\Lambda}, \Omega_{M}, h] = [0.714, 0.286, 0.696]$), we calculated the photometric distance to the burst source, $D_{L} = 1.82 \ast 10^{28}$ cm, and the equivalent isotropic energy emitted in the gamma-ray range (1, 10000) keV, $E_{iso} = (5.24 \pm 0.18) \ast 10^{52}$ erg. On Amati diagram GRB 090510 lies above the main correlation region (Fig. 8), which confirms its association with the class of short bursts. The parameter $E_{iso}$ for the precursor candidate can also be estimated using a power law with a break (Band) model. It is $E_{iso} = (7.7 \pm 1.9) \ast 10^{50}$ erg. On Amati diagram the precursor candidate also lies above the main correlation region, i.e., it occupies a position typical for short GRBs (Fig. 8).

The results of our spectral analysis for GRB 090510 and our estimate of the parameter $E_{iso}$ for the main phase are consistent with the results obtained previously (Ackermann et al. 2010; Gruber et al. 2014; Muccino et al. 2013).

Thus, our detailed study of GRB 090510 showed that it belongs to the class of short bursts. However, no convincing evidence that the precursor candidate has a different nature/emission mechanism compared to the main GRB phase has been found.

\subsection*{The Averaged Light Curve}

To construct the combined light curve, we modified our complete sample of 519 GRBs with a duration $T_{90}$ less than 2 s.

The GRB light curves usually have a fairly complex structure consisting of a superposition of FRED (fast rise - exponential decay) pulses; therefore, their observed duration depends on the statistical significance above the background level. Using BATSE GRBs as an sample, Koshut et al. (1996) showed that the observed duration $T_{90}$ systematically dropped relative to the true value with decreasing statistical significance of the burst starting from S/N $\sim 10~\sigma$. This means that a long burst with a true duration $T_{90}$ > 2 s at a low statistical significance, S/N $\sim 6~\sigma $ (the lower significance limit for the bursts of our complete sample), can have a duration $T_{90}$ < 2 s and, hence, can be erroneously classified as a short one (see also the next section and Fig. 11). To minimize the number of such events, we excluded the bursts with a statistical significance of less than ten standard deviations from our complete sample. In addition, we excluded all bursts with precursor candidates.

The averaged light curve is the sum of the individual light curves aligned relative to the peak flux of the main phase on a time scale of 0.05 s (for details, see the “Data Processing Algorithm” Section). We have no a priori information about the presumed regular precursor (its duration, location relative to the main burst phase, intensity, etc.). Therefore, we made a “blind” search for a possible regular precursor in the averaged light curve.

We found no significant (more than three standard deviations) excess on any time scale from 0.05 (the time resolution of the original data) to 48 s (the longest investigated interval before the burst onset) in the averaged light curve in the interval (-50, -2) s, suggesting the absence of a regular precursor for the short GRBs of our sample. Obviously, the position of the possible regular precursor relative to the peak of the main GRB phase can vary. In this case, when the light curves of various bursts are summed, the intensities of the precursors of the individual light curves will be averaged. As a consequence, the possible regular precursor will be “smeared” over the entire averaged light curve.

We can estimate an upper limit on the relative intensity of the regular precursor by assuming it to be a positive deviation with a 3 $\sigma$ significance. Our estimate of the maximum intensity of the presumed regular precursor will not depend on its duration but will depend only on the length of the interval being investigated. We will use the same intervals as above for our estimation, namely (-5, -2) and (-50, -2) s relative to the peak of the averaged light curve.

The fluence for the main phase of the averaged light curve is $F = 4.78 \ast 10^6$ counts. The fluence corresponding to a positive background fluctuation in the averaged light curve with a statistical significance of 3 standard deviations in the interval (-5, -2) s is $B = 3.81 \ast 10^4$ counts. Hence we will find that the presumed regular precursor is weaker than the main phase at least by a factor of $F / B \simeq 125$. A similar estimation for the interval (-50, -2) s gives a ratio $F / B \simeq 30$. The latter estimate is most conservative, provided that the precursor (of any duration) is in the interval (-50, -2) s. Interestingly, the above intensity ratios are larger than those for the main burst phase and the precursor candidate for all the previously considered bursts: 26, 3.6, and 29 for GRB 090510, GRB 100717, and GRB 130310, respectively (see Table 2).

\section*{DISCUSSION}

\subsection*{The Bimodal Duration Distribution of GRBs and the Problem of Their Classification}

The most complete sample of GRBs in the SPI-ACS experiment is contained in Savchenko et al. (2012). Therefore, in our analysis we used the distribution of GRBs in duration $T_{90}$ constructed in Savchenko et al. (2012). It is presented in Fig. 11. To fit this distribution, we used log-normal functions (below referred to as model curves) corresponding to the modes of long and short bursts (indicated in Fig. 11 by the thin smooth curves). The combined model curve is indicated by the smooth thick line. The maxima of the modes of long and short bursts correspond to a duration of 20.6 and 0.43 s, respectively. The fraction of short GRBs among all bursts calculated as the ratio of the areas under the model curve for the mode of short bursts and the combined model curve is 22\%. Interestingly, the FWHMs (full widths at half maximum) of the model curves for long and short bursts coincide within the 1 $\sigma$ error limits.

\begin{figure}[h]
\epsfxsize=16cm \hspace{-0.3cm}\epsffile{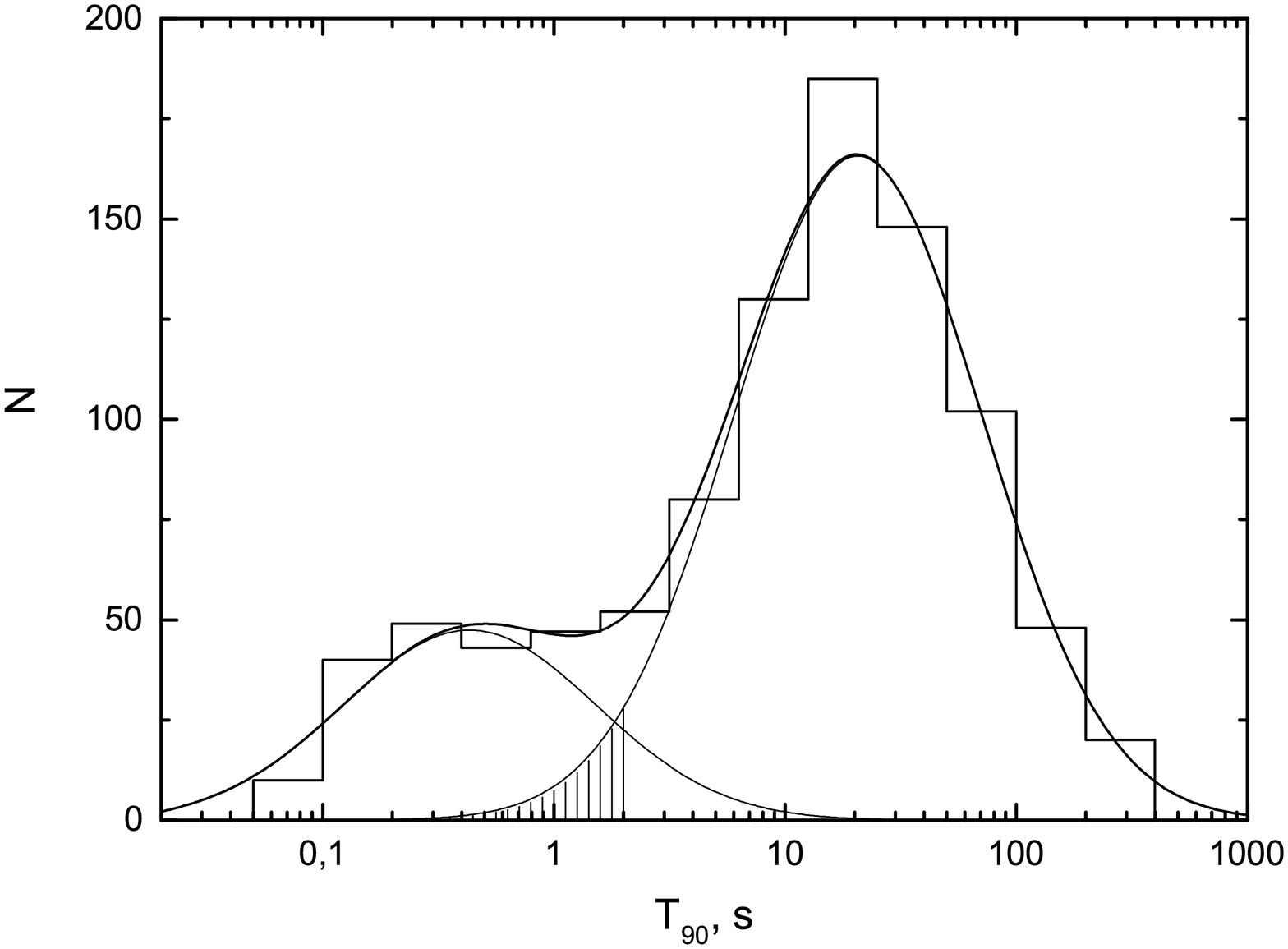}
\\ \textbf{Fig. 11.} \emph{Duration distribution of GRBs detected in the SPI-ACS/INTEGRAL experiment constructed from the data of Savchenko et al. (2012). The GRB duration $T_{90}$ in seconds is along the horizontal axis; the number of bursts with a give duration is along the vertical axis. The smooth thin curves indicate the fit to the distribution by two log-normal functions; the smooth thick curve indicates the sum of these functions. The hatched part of the log-normal model curve for long GRBs shows the fraction of these events among the GRBs with a duration of less than 2 s.}
\end{figure}

The model curves intersect at 1.85 s. This value is very close to what is used as the boundary one in the classification of GRBs by duration: GRBs with a duration of less than 2 s are usually classified as long ones. However, the position of this point also depends on the energy range in which a particular experiment operates (see, e.g., Minaev et al. 2010b). It follows from Fig. 11 that not all long burst have a duration of more than 2 s: the model curve extends to $\sim$ 0.5 s (the corresponding region of the model curve is hatched). At 1.85 s the probabilities that a specific GRB is long or short are equal. If the duration of a GRB is less than 2 s, then the probability that it is long is lower than the probability that it is short, but, at the same time, it is nonzero. Thus, the fraction of long GRBs among the GRBs with a duration of less than 2 s is $\simeq$ 9\% (the hatched region in Fig. 11). This means that about 50 of our investigated 519 GRBs with a duration of less than 2 s can be representatives of the class of long bursts. At the same time, the fraction of representatives of short bursts in the sample of bursts with a duration of more than 2 s is 2.5\%.

However, for each specific case we cannot ascertain whether it is a long burst without carrying out additional studies (for example, studies of the energy spectrum and its evolution, the spectral lag, the X-ray and optical afterglows, the signatures of a supernova, etc.).

The fraction of long GRBs with precursors varies within the range from 3\% to 20\% in the BATSE experiment (Koshut et al. 1995; Lazzati 2005). This means that among all of our investigated GRBs with a duration of less than 2 s, about 50 can be long ones, and, in turn, from 1 to 10 of them can have precursors (assuming the SPI-ACS and BATSE experiments to be identical).

We found precursor candidates for three GRBs, one of which, GRB 071030, is insufficiently significant. GRB 100717 is close in its properties to the class of long bursts. GRB 130310 exhibits the properties of both long and short bursts; therefore, it may well belong to the class of long GRBs (including the subclass of long bursts with precursors). Despite the fact that the properties of GRB 090510 are typical for “classical” short bursts, it can also be a long burst belonging to the hatched part of the model curve in Fig. 11, because its duration $T_{90} = 0.98 \pm 0.07$ s.

Thus, we found no significant candidates for the subclass of short GRBs with precursors. We can estimate an upper limit on the number of short GRBs with precursors in the SPI-ACS/INTEGRAL experiment. If GRB 090510 and GRB 130310 are assumed to be actually short GRBs with precursors, then there are no more than $\sim 0.4\%$ of such events in the SPI-ACS experiment among all short GRBs.

\subsection*{Precursor Candidate Selection Criteria and the Results of Other Works}

So far there is no single definition of precursor. In Troja et al. (2010) the less intense and shorter burst activity episode preceding the main one is considered to be a precursor. In Koshut et al. (1995) an additional condition is imposed on the precursor properties: the time interval between the precursor and the main GRB episode must exceed the duration Т$_{90}$ of the main episode. Since the GRB light curves generally have a complex structure and consist of several pulses, both overlapping between themselves and well separated in time, the possibility that the component of the light curve classified as a precursor is actually the component of the main GRB phase must not be ruled out. The probability of this depends largely on the rigidity of the precursor candidate selection criterion. The mildest selection criterion is in Troja et al. (2010), where a constraint is imposed only on the relative duration and intensity of the precursor.

GRB 090510 and GRB 100717 are typical examples. GRB 090510 was classified (Troja et al. 2010) as a short burst with a precursor. In this paper we found no convincing evidence that the precursor candidate has a different nature compared to the main GRB phase. The same is also true for GRB 100717, whose precursor candidate is apparently the beginning of the main burst phase.

Therefore, it is necessary to minimize the probability that the detected precursor candidate is the beginning of the main episode of a short GRB using a more rigid precursor candidate selection criterion. Since the hypothetical precursor presumably has a different nature, its spectral–temporal properties can differ from those of the main burst phase. It can also be offset from the main phase by a considerable time interval. In most cases, the duration of the main phase of a short GRB does not exceed 2 s. Hence we proposed the following precursor selection criterion: the time interval between the precursor and the main episode must exceed 2 s, and its properties must differ from the properties of the main episode. These properties include at least the energy spectrum, the variability, and the shape and number of pulses in the light curve. In a similar situation with long GRBs, in the case with GRB 160625B, the spectral properties of the precursor differ significantly from those of the main phase (Zhang et al. 2017).

Although our criterion (the time interval between the precursor candidate and the beginning of the main burst phase must exceed 2 s) allows the probability that the precursor candidate is actually the beginning of the main burst phase to be minimized, the precursor candidate that we found for GRB 100717 turned out to be the beginning of the main burst phase. However, it should be emphasized that GRB 100717 is an exceptional example: it belongs to the class of long bursts whose characteristic time scales (the duration, the time interval between the individual structures of the main burst phase) are a factor of $\sim$ 50 greater than those for short bursts (see Fig. 11). Therefore, the precursor candidate selection criterion for long bursts must be different.

Finally, the search for precursors and the mutual study of their parameters for any GRBs (short and long) detected in various experiments (in our case, SPI-ACS and GBM) make it possible to compare and calibrate the SPI-ACS sensitivity to short low intensity bursts, which is necessary in searching for the short GRBs that may accompany the sources of gravitational waves (Connaughton et al. 2016; Savchenko et al. 2016).
\\

\subsection*{CONCLUSIONS}

We analyzed both the individual light curves for our sample of 519 short GRBs and the averaged light curve of 372 brightest short bursts from this sample.

The following precursor candidate selection criterion was used: the time interval between the main GRB phase and the precursor candidate must be within the ranges (-5, -2) and (-50, -2) s on time scales of 0.1 and 5 s, respectively; the statistical significance of the candidate must be greater than six standard deviations.

With this criterion we found three candidates for GRB 071030, GRB 100717, and GRB 130310. The precursor candidate for GRB 071030 is insufficiently reliable; the candidates for GRB 100717 and GRB 130310 were confirmed in the GBM/Fermi experiment and were studied in detail. The precursor candidate for GRB 090510 from Troja et al. (2010) was investigated additionally.

It was shown that GRB 100717 could be a representative of the class of long bursts. For the remaining two precursor candidates of GRB 130310 and GRB 090510 we found no convincing evidence that they are precursors and have a different nature/emission mechanism compared to the main burst episode. Moreover, one might expect approximately the same number of precursors from the fraction of long GRBs erroneously classified as short ones based only on the duration selection criterion. Therefore, an unequivocal conclusion of whether the precursor candidates found are the precursors of precisely short GRBs cannot be reached.

On the other hand, the combined light curve contains no signatures of a regular precursor. At the most conservative estimate, the possible regular precursor is weaker than the main phase of a short burst by more than a factor of 30.

Thus, the precursors of short GRBs detected in the SPI-ACS/INTEGRAL experiment, if any, exist only in a small number of events, more specifically, for less than 0.4\% of the short GRBs.

 \noindent

 \pagebreak

\clearpage

\end{document}